# Granular Matter and the Marginal Rigidity State.

R Blumenfeld*†, SF Edwards† and RC Ball*,

*Department of Physics, University of Warwick, CV4 7AL Coventry, UK.

†Polymers and Colloids, Cavendish Laboratory, Madingley Road,

Cambridge CB3 0HE, UK.

**Model experiments are reported on the build-up of granular piles in two dimensions. These show that as the initial density of falling grains is increased, the resulting pile has *decreasing* final density and its coordination number approaches the low value predicted for the theoretical marginal rigidity state. This provides the first direct experimental evidence for this state of granular matter. We trace the decrease in the coordination number to the dynamics within an advancing yield front between the consolidated pile and the falling grains. We show that the front's size increases with initial density, diverging as the marginal rigidity state is approached.**

Recent theoretical works[1-5] have suggested that a distinctive, Marginal Rigidity State exists for rigid cohesionless grains. This state is characterised by low connectivity and exhibits stress transmission governed by its geometry alone. It has yet to be established whether this applies to real granular matter, such as sand or grain, but it is a candidate to underpin recent success in modelling macroscopic stress transmission[6,7]. Here we present the first direct experimental evidence, from idealised granular systems in two dimensions as shown in figure 1, that the marginal solid state of matter does indeed exist. Our experiments also reveal a diverging lengthscale, suggesting the state is a critical phenomenon.

The key idea behind this state is that rigid grains falling to form a pile will rearrange and consolidate, until their mean coordination number z reaches a critical value $z_c$ at which the net force and torque on each grain can first be balanced by the intergranular forces. At this Marginal Rigidity State the forces, and hence the transmission of stress, are exactly determined by balance alone, without reference to any internal constitutive behaviour of the individual grains. This state separates fluid from conventional solid: so long as $z < z_c$, mechanical balance cannot in general be obtained and the system is fluid. By contrast when $z > z_c$ the system can be mechanically balanced, but additional constraints are needed to determine the intergranular forces. The extra constraints usually come from constitutive stress-strain relations and consistency of the strain field, leading to stress transmission typical of traditional solids.

For convex grains with friction the critical coordination number predicted is notably low, $z_c = 3$ and 4 in two and three dimensions respectively[4]. The corresponding values for sequential packing are 4 and 6, and regular sphere close packing gives even higher values of 6 and 12 respectively. Therefore, the coordination number is a key feature to test in order to identify the Marginal Rigidity State.

Our experiments are illustrated in Figure 1. In an analogue of free fall, grains were conveyed by the moving base until they encountered either the collecting boundary or its accumulating pile. Then the force of sliding friction relative to the moving base played a role analogous to gravitational force on a stationary mass. Figure 2a shows how the final density $\rho_p$ of the resulting pile *decreases* with increasing the initial density $\rho_i$ at which grains were fed in. We believe this decrease may be characteristic of a jamming system, and our discussion of the front below offers some insight as to how this comes about. A good empirical fit to the density given by

$$\rho_p = 0.446 + 0.072\, e^{-2.7\, \rho_i} \quad (1)$$

and the extrapolated value at which the two densities match is given by $\rho_c = 0.465(5)$. This latter is the natural limiting density of our experimental series.

Figure 2b shows how the coordination number z varies with initial density $\rho_i$ in our experiments. The striking feature is that as $\rho_i$ increases, z decreases towards an extrapolated value $z(\rho_c) = 3.1$, in qualitative agreement with the simple prediction $z_c = 3$ of marginal rigidity. We know of no other interpretation of how starting at higher density leads to lower coordination number of this value.

Due to their shape, our grains are just capable of forming double contacts, a complication that is not addressed by the theory for convex grains. We have counted these as two contacts each, whereas a double contact imposes only three constraints in total rather than four for two separate contacts: one constraint is thus 'wasted' per double contact. Counting coordination number as we have done, this means that the theoretically predicted critical coordination number $z_c$ should be shifted (in two dimensions) to

$$z_c = 3 + w_2 \quad (2)$$

where $w_2$ is (without any double counting) the number of double contacts per grain. From our experiments we measured $w_2 \approx 0.1$ with no significant dependence on $\rho_i$, leading to quantitative agreement of $z_c$ with the extrapolated value $z(\rho_c)$.

At lower initial densities the measured coordination number rises with decreasing initial density, approaching $z_0 = 3.7$ as $\rho_i$ tends to zero. This is in fair agreement with the value 4 expected for sequential packing, where each arriving grain would come fully to rest with two contacts before another encounters the pile. Another way to view this would be that since each grain encounters an already rigid pile, one constraint is wasted out of the four from two contacts, in precise analogy to our discussion of double contacts above. In the experiments, we observe that occasionally an incoming grain will dislodge other previously stationary grains before settling down. Thus the positioning is not strictly sequential and this explains why the coordination number does not reach as high as 4.

At intermediate densities, the coordination numbers can be interpreted in terms of intermediate levels of constraint wastage. A constraint is wasted every time a new loop of contacts is closed where the corresponding jaw has only one hinge. Every wasted constraint introduces an ambiguity regarding the transmission of forces in the pile. A double contact is special in this respect, because its force ambiguity is localised within the closed doublet of particles and therefore does not affect macroscopic stress transmission. Precisely which wasted constraints preserve marginal rigidity in this way is an open question.

While the above results verify the existence of the Marginal Rigidity State, our experiments can also be interpreted on a macroscopic level in terms of a yield front that propagates ahead of the consolidated pile. We define the front, at any given moment, as all the grains that have already collided with the pile (and its connected front) but have not yet reached their final position in the consolidated configuration. Figure 1b shows three stages in the time evolution of a pile with its front coloured red. Figure 3a shows the front for three different values of $\rho_i$. It can be observed that the higher the initial density the deeper the front. In the sequential packing regime $\rho_i \rightarrow 0$, the front is less than a monolayer and each falling grain collides with an almost completely rigid pile, hence the significant constraint wastage and higher coordination number. At higher $\rho_i$, falling grains encounter a layer that is still rearranging: then the constraints from their contacts are less likely to be wasted, resulting in a lower final coordination number.

We characterised the front by its mass per unit width of the base, m, whose increase with $\rho_i$ can be seen in the main plot of Figure 3b. On the basis of this data and a simple theory presented below, we conjecture that the front size in an infinite system diverges as $\Delta\rho = \rho_P - \rho_i \rightarrow 0$. In our experiments the front size approaches our (limited) system size by $\rho_i = 0.3$.

We can estimate theoretically the divergence as $\Delta\rho \rightarrow 0$ of the front depth $\xi$ by considering the time available for grains to rearrange whilst in the front. From conservation of grains it follows that the mean velocity of grains *relative to the moving front* is approximately $v_r = v_0(\rho_p+\rho_i)/(2\Delta\rho)$, where $v_0$ is the speed of the falling grains relative to the pile. The *time* a grain resides in the front is then estimated by $\tau \approx \xi/v_r$. In this time the maximum sideways displacement that a grain can achieve is of order $\Delta r \approx \tau v_0$, in terms of which we can now express the front width as

$$\xi \approx \Delta r (\rho_p+\rho_i)/(2\Delta\rho) \qquad (3).$$

We would expect $\Delta r$ to be of order a grain size for reorganisation of the grain distribution to be achieved from the feed into the pile, and certainly by experimental observation we are confident that if $\Delta r \to 0$ as $\Delta\rho \to 0$, it does so only very slowly. Hence we are led to the front width diverging as

$$\xi \propto \Delta\rho^{-\nu} \qquad (4)$$

with $\nu=1$ if $\Delta r$ is set by the grain size.

The very existence of a diverging front size as the marginally rigid state is approached strongly suggests that the latter is a continuous critical point, analogous to various self-organising critical systems. If so the exponent $\nu$ should be insensitive to details of the grains, corresponding to a new universality class, but it should be sensitive to the two dimensional nature of our experiments and to the dominance of friction over inertia and hydrodynamics.

We believe that three dimensional experiments would give similar results, and the classic experiment of Bernal and Mason[8] should be noted. They have found that loose packing of ball bearings within a rolled cylinder with dimpled sides gave $z = 5.5$, a value which is intermediate between the Marginal Rigidity values $z_c = 4$ with friction and $z_c = 6$ for frictionless spheres, and below that of sequential packing, $z=6$. While to measure coordination number *in situ* in three dimensions is harder, the front depth may be easier to measure via the density profile.

**Methods**

Model noncircular grains of approximately $1.5\pm0.2\text{cm}^2$ were cut from $1.6\pm0.2\text{mm}$ cardboard using a stainless steel punch. The fibrous nature of the cardboard ensured that the cut edges gave high coefficient of intergranular friction, which we estimated from slipping tests to be at least 50. The experiment was carried out on a horizontal glass plate, with the pile building up within a ⊔-shaped collector surface of similar material (see figure 1) and internal dimensions approximately $18\times24\text{cm}^2$. The grains were initially placed on a thin transparent film lying on top of the plate at a notionally random distribution subject to the requirement of no contacts and reasonably uniform density.

The 'free falling' grains were conveyed towards the collector at an approximate speed of 0.5m/min (maximum): at high initial densities this was done by moving the collector towards the grains. At low initial densities the film was moved towards the stationary collector, and the 'initial' distribution of grains was created only as it approached the pile. Winding mechanisms proved vulnerable to stick-slip motion of the film and were rejected in favour of driving the experiment directly by hand: care was taken to maintain a constant advance rate and particularly to avoid relative transverse motion between the collector surface and the free falling grains. The slow fall rate ensured both negligible grain deformation and minimisation of inertial effects.

The data was accumulated in the form of photographs of the growing pile, taken at regular intervals during the process, and a photograph of the consolidated final pile. For each final pile we have counted its grains, the boundary grains, the contacts and the double contacts. Because the theoretical predictions of coordination number are based fundamentally on the expectation of three constraints per grain at the marginal state, we measured coordination number as 2x(total number of contacts)/(total number of grains), which, if calculating coordination from individual grains, means double counting coordinations from a grain to the boundary. The (relatively few) double contacts were counted as two separate contacts.

The photographs taken during each piling process were used to identify the yield front by comparing the intermediate positions of the grains to their position in the final pile. The comparison was done by superposing the two photographs on top of a lightbox and determining overlaps by eye.

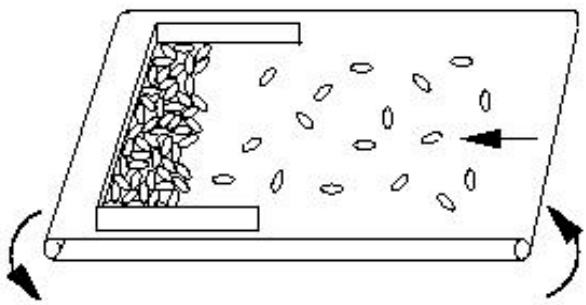 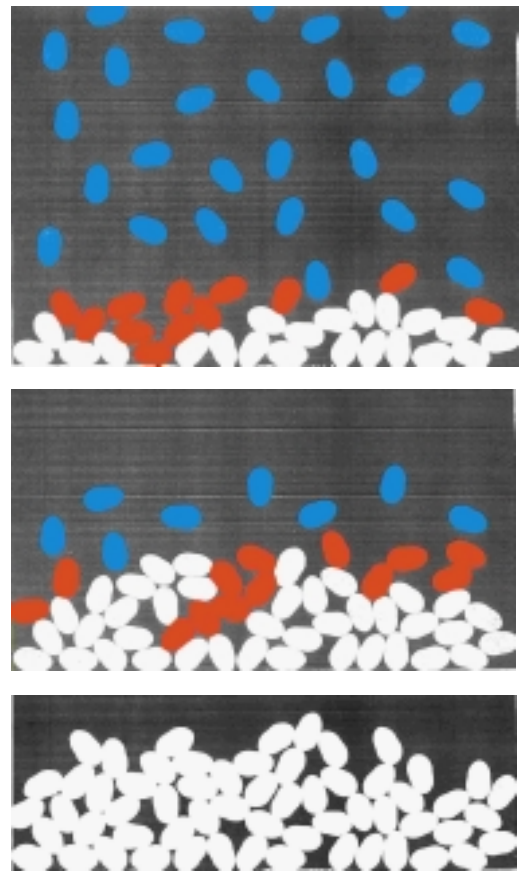

**a**  **b**

Figure 1 The experiment. **a**, Sketch of our experimental setup: cardboard grains are conveyed by a moving surface towards a stationary ⊔–shaped collector of similar material. The grains are effectively rigid, with high mutual friction, and the slow advance rate minimises inertial effects. **b**, The growth of a pile, with time running from top image to bottom. The grains coloured blue are in an analogue of free free fall, comoving with the base towards the growing pile. The white grains have come permanently to rest, and friction relative to the base supplies an analogue of gravitational force. The grains coloured red have not fully consolidated (see also figure 3).

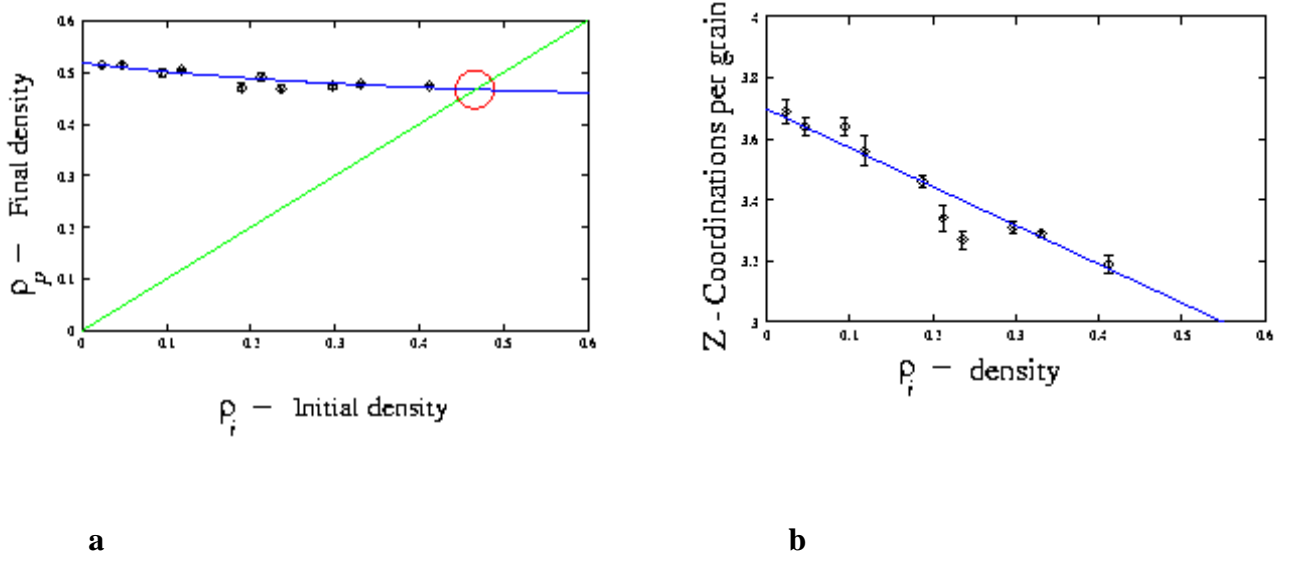

**a**  **b**

Figure 2 The behaviour of the consolidated pile. **a**, The final density of the consolidated pile, $\rho_p$, as a function of the initial density of the free falling grains, $\rho_i$. We suggest the decreasing behaviour is characteristic of the jamming nature of our system. The line interpolating the experimental points is the numerical fit given in eq. (1), and the circled point is where the initial and final densities extrapolate to equal, at density $\rho_c = 0.465(5)$, which is the natural limiting density of our experiments. **b**, The mean coordination number z of each consolidated pile as a function of the initial density, $\rho_i$. The data extrapolate to z =3.1 at the limiting density $\rho_c$, in good agreement with prediction for the marginal rigidity state. The coordination number approaches z=3.7 in the zero density limit, which approximates to sequential packing. In both figures 2 and 3 the error bars reflect one standard error in the mean from several repetitions of the experiment.

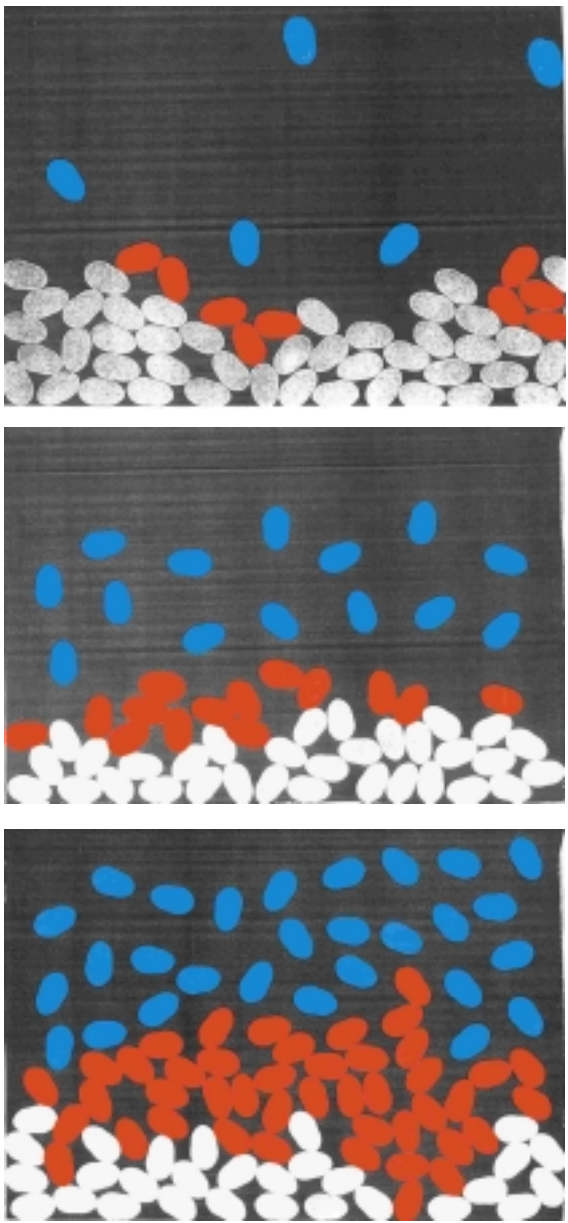

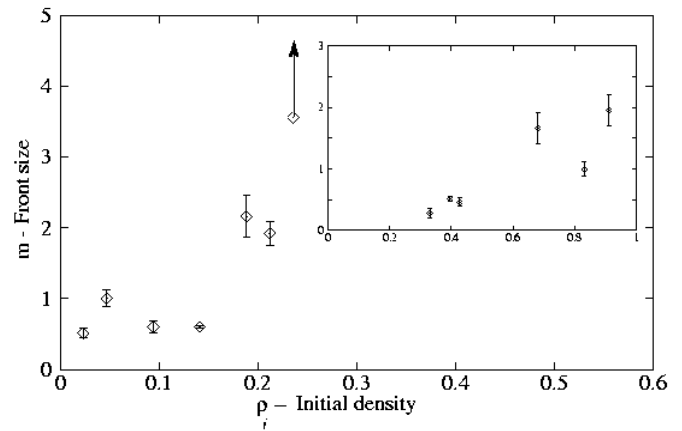

a                                                          b

Figure 3 The behaviour of the front. **a,** The front (coloured red) of grains which have encountered the pile but not yet reached their final consolidated positions, from three different experiments at initial densities 0.024 (top), 0.142 (middle), and 0.236 (bottom). At the low density the front is less than one grain deep while at the high density it is more than half the pile and the consolidation is highly cooperative. **b,** The measured front mass m, in grains per unit of base, as a function of the initial density. This shows a diverging trend, and the point at $\rho=0.236$ should be regarded a lower bound. The inset shows $1/m$ plotted against $(\rho_p - \rho_i)/(\rho_p + \rho_i)$, which a simple analysis (see text) would predict to be a straight line through the origin corresponding to a critical divergence $m \propto \Delta\rho^{-1}$.